\documentclass{osa-article}

\journal{oe}

\articletype{Research Article}

\begin{document}

\title{Simple and efficient way to generate superbunching pseudothermal light}

\author{ Jianbin Liu,\authormark{1, 2,*}  Rui Zhuang, \authormark{1}Xuexing Zhang, \authormark{1} Chaoqi Wei,\authormark{1} Huaibin Zheng, \authormark{1}  Yu Zhou, \authormark{3} Hui Chen, \authormark{1} Yuchen He, \authormark{1} and Zhuo Xu \authormark{1}}

\address{\authormark{1}Electronic Materials Research Laboratory, Key Laboratory of the Ministry of Education \& International Center for Dielectric Research, School of Electronic Science and Engineering, Xi’an Jiaotong University, Xi'an 710049, China\\
\authormark{2}The Key Laboratory of Weak Light Nonlinear Photonics (Nankai University, Tianjin 300457), Ministry of Education, China\\
\authormark{3}MOE Key Laboratory for Nonequilibrium Synthesis and Modulation of Condensed Matter, Department of Applied Physics, Xi’an Jiaotong University, Xi’an, Shaanxi 710049, China}

\email{\authormark{*}liujianbin@xjtu.edu.cn} 



\begin{abstract}
By modulating the intensity of laser light before the rotating groundglass,  the well-known pseudothermal light source can be modified into superbunching pseudothermal light source, in which the degree of second-order coherence of the scattered light is larger than 2. With the modulated intensities following binary distribution, we experimentally observed the degree of second- and third-order coherence equaling 20.45 and 227.07, which is much larger than the value of thermal or pseudothermal light, 2 and 6, respectively. Numerical simulation predicts that the degree of second-order coherence can be further improved by tuning the parameters of binary distribution. It is also predicted that the quality of temporal ghost imaging can be improved with this superbunching pseudothermal light. This simple and efficient superbunching pseudothermal light source provides an interesting alternative to study the second- and higher-order interference of light in these scenarios where thermal or pseudothermal light source were employed.
\end{abstract}



\section{Introduction}\label{introduction}
Since its invention \cite{martienssen}, pseudothermal light has played an essential role in the development of second- and higher-order interference experiments in quantum optics, such as  ghost imaging \cite{gi-2004,gi-2005,gi-2009,gi-2015}, two-photon interference \cite{ti-2005,ti-2006},  intensity interferometer \cite{ii-2010,ii-2018}, quantum erasers \cite{dc-2014}, and so on. Pseudothermal light was generated by inputing single-mode continuous-wave laser light onto a rotating groundglass \cite{martienssen} or into a multimode fiber \cite{apb-2017}. The properties of the generated pseudothermal light are similar as the ones of true thermal light except the degeneracy parameter and coherence time are tunable \cite{martienssen}, which makes the second- and higher-order interference of light much easier with pseudothermal light than the ones with true thermal light \cite{hbt-book}. 

Recently, we proposed a new type of light source called superbunching pseudothermal light source by employing laser light, pinholes, and multiple rotating groundglasses \cite{zhou-2017, bai-2017}. The degree of second-order coherence, $g^{(2)}(0)$, of superbunching pseudothermal light can be much larger than the one of pseudothermal or thermal light, 2. For instance, we experimentally observed $g^{(2)}(0)$ equaling 7.10  by employing three rotating groundglasses, which is 3.5 times of the one of thermal or pseudothermal light. It is predicted that $g^{(2)}(0)$ of superbunching pseudothermal light equals $2^N$ if $N$ rotating groundglasses were employed, where $N$ is a positive integer \cite{zhou-2017}. However, the problem with adding more rotating groundglasses to create superbunching pseudothermal light is that the uncertainty of the obtained $g^{(2)}(0)$ can be very large if more than three rotating groundglasses were employed. If this type of superbunching pseudothermal light is employed in temporal ghost imaging \cite{tgi-2016}, the visibility can indeed be increased by adding more rotating groundglasses. However, the quality of the retrieved image may actually be decreased. For instance, the visibility of temporal ghost imaging can be increased from 4.7\% to 75\% if six rotating groundglasses were employed instead of one \cite{liu-2018}. However, the signal-to-noise ratio of the retrieved image decreases dramatically as the number of rotating groundglasses increases, which can be clearly seen in the results shown in Figs. 3(a) - 3(f) in Ref. \cite{liu-2018}.

In order to solve the problem that the uncertainty of $g^{(2)}(0)$ increases as the value of $g^{(2)}(0)$ increases in superbunching pseudothermal light with multiple rotating groundglasses \cite{zhou-2017}, we employed an electromagnetic optical modulator (EOM) to modulate the intensity of the incident laser light before the rotating groundglass. Several typical signals, such as sinusoidal,  triangle, and white noise, were employed to drive EOM to observe two-photon superbunching effect, in which the maximal observed $g^{(2)}(0)$ equals 3.32 \cite{zhou-2019}. In a recent study, Straka \textit{et al.} employed an acousto-optical modulator to modulate the intensity of laser light to generate arbitrary classical photon statistics and observe very large value of $g^{(2)}(0)$. However, lengthy mathematical calculations are required in their experiments  \cite{straka}. In this paper, we will combine the methods above \cite{zhou-2019, straka} to find a simple and efficient way to generate superbunching pseudothermal light with tunable degree of second-order coherence. $g^{(2)}(0)$ equaling 20.45 and $g^{(3)}(0)$ equaling 227.07 are experimentally observed, where $g^{(3)}(0)$ is the degree of third-order coherence. Theoretically, the degree of second- and higher-order coherence can be further increased. More importantly, the uncertainty of $g^{(2)}(0)$ does not increases as the value of $g^{(2)}(0)$ increases as the one in multiple rotating groundglasses case \cite{zhou-2019}, which makes it a perfect light source to obtain high quality temporal ghost imaging \cite{tgi-2016}.

The rest of the paper is organized as follows. Employing two-photon interference theory to calculate the second-order coherence function of superbunching pseduothermal light is in Sect. \ref{theory}. Section \ref{experiments}  includes the experimental observation of superbunching pseudothermal light with EOM and rotating groudglass. Numerical simulations and discussions are in Sect. \ref{discussion}. Our conclusions are summarized in Sect. \ref{conclusion}.

\section{Theory}\label{theory}

Both quantum and classical theories can be employed to calculate the second-order interference of classical light \cite{glauber,glauber1,sudarshan}. We will employ two-photon interference in Feynman\rq{}s path integral theory to calculate the second-order coherence function of superbunching pseudothermal light \cite{feynman-p}, since it is easier to understand the physics of the mathematical calculations \cite{liu-2010}. The second-order coherence function of superbunching pseudothermal light with EOM and rotating groundglass in a Hanbury Brown-Twiss (HBT) interferometer can be expressed as  \cite{liu-cpb,zhou-2019,hbt,hbt-1}
\begin{equation}\label{g2-1}
G^{(2)}({\bf{r}_1},t_1;{\bf{r}_2},t_2)=\langle |\sqrt{P_{a1}P_{b2}}A_{a1b2}+\sqrt{P_{a2}P_{b1}}A_{a2b1}|^2\rangle,
\end{equation}
where $({\bf{r}_1},t_1)$ and $({\bf{r}_2},t_2)$ are the space-time coordinates of photon detection events at D$_1$ and D$_2$, respectively. D$_1$ and D$_2$ are two single-photon detectors in a HBT interferometer. $P_{\alpha\beta}$ and $A_{\alpha\beta}$ are the probability and probability amplitude of photon $\alpha$ detected by D$_\beta$, respectively ($\alpha=a$ and $b$, $\beta=1$ and 2). $\langle ... \rangle$ represents ensemble average. $A_{a1b2}$ is a two-photon probability amplitude, which equals the product of two single-photon probability amplitudes, $A_{a1}$ and $A_{b2}$ \cite{feynman-p}. The single-photon probability amplitude $A_{\alpha\beta}$ is
\begin{equation}\label{a}
A_{\alpha\beta}=e^{i\varphi_\alpha}K_{\alpha\beta},
\end{equation}
in which $\varphi_\alpha$ is the initial phase of photon $\alpha$, and $K_{\alpha\beta}$ is the Feynman\rq{}s photon propagator that describes photon $\alpha$ traveling to D$_\beta$. If a point light source is employed, Feynman\rq{}s photon propagator \cite{liu-2010, peskin},
\begin{equation}\label{k}
K_{\alpha\beta}=\frac{\exp[-i({\bf{k}_{\alpha\beta}\cdot r_{\alpha\beta}}-\omega_\alpha t_\beta)]}{ r_{\alpha\beta}},
\end{equation}
is the same as Green function for point light source in classical optics \cite{wolf}, in which $\bf{k}_{\alpha\beta}$ and $\bf{r_{\alpha\beta}}$ are the wave and position vectors of the photon $\alpha$ being detected at D$_\beta$, respectively. $r_{\alpha\beta}$ equals $|\bf{r}_{\alpha\beta}|$. $\omega_{\alpha}$ and $t_{\beta}$ are the frequency and time for the photon $\alpha$ being detected at D$_\beta$, respectively.

Substituting Eqs. (\ref{a}) and (\ref{k}) into Eq. (\ref{g2-1}) and ignoring the spatial parts, the second-order temporal coherence function can be simplified as \cite{zhou-2019},
\begin{eqnarray}\label{G2t}
&&G^{(2)}(t_1,t_2)\nonumber\\
 & \propto & \langle P_{a1}P_{b2}  \rangle + \langle P_{a2}P_{b1}\rangle \nonumber\\
&&+2\langle \sqrt{P_{a1}P_{b2}P_{a2}P_{b1}} \rangle \text{sinc}^2\frac{\Delta \omega (t_1-t_2)}{2}, \\
&\propto & \Gamma(t_1-t_2) [1+ \text{sinc}^2\frac{\Delta \omega (t_1-t_2)}{2}],\nonumber
\end{eqnarray}
where $\Gamma(t_1-t_2)$ is the intensity correlation function of laser light before rotating groundglass, and $\Delta \omega$ is the frequency bandwidth of pseudothermal light caused by the rotating groundglass. The correlation expressed by Eq. (\ref{G2t}) can be divided into two parts. One part of intensity correlation is generated by rotating groundglass and expressed by $1+\text{sinc}^2[{\Delta \omega (t_1-t_2)}/{2}]$, which is a typical second-order temporal coherence function of thermal or pseudothermal light \cite{shih-book}. The other part is caused by the intensity modulation of EOM, which is expressed as $\Gamma(t_1-t_2)$. If there is no intensity modulation through EOM, $\Gamma(t_1-t_2)$ equals 1 and Eq. (\ref{G2t}) is simplified as the second-order temporal coherence function of thermal or pseudothermal light \cite{shih-book}. In order to obtain larger value of $g^{(2)}(0)$ of superbunching pseudothermal light, one needs to increase the intensity correlation generated by EOM. Here, we will employ binary distribution for example to modulate the intensity of laser light, in which the correlation can be varied by tuning the corresponding parameters. 

\section{Experiments}\label{experiments}
\begin{figure}[htbp]
\centering
\includegraphics[width=100mm]{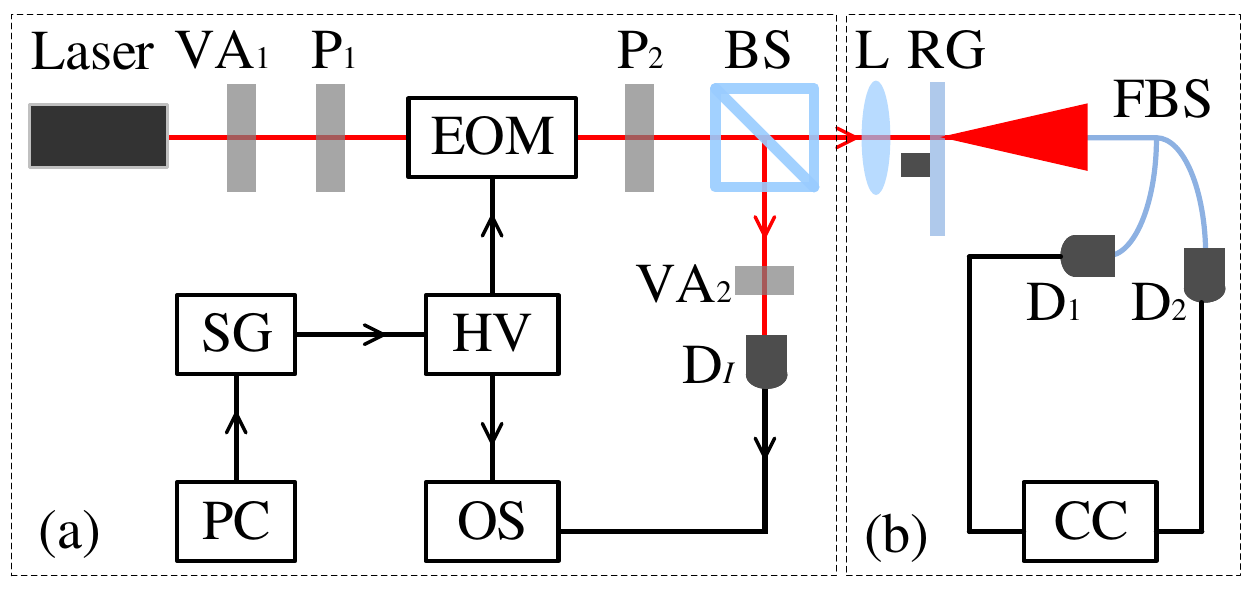}
\caption{Experimental setup to observe two-photon superbunching effect. Laser: single-mode continuous-wave laser. VA: variable attenuator. P: polarizer. PC: computer. SG: signal generator. EOM: electro-optical modulator. HV: high-voltage amplifier. OS: oscilloscope. BS: 1:1 non-polarizing beam splitter. L: lens. RG: rotating groundglass. FBS: 1:1 single-mode fiber non-polarizing beam splitter. D$_I$: intensity detector. D$_1$ and D$_2$: single-photon detectors. CC: two-photon coincidence count detection system. See text for detail descriptions.}\label{1}
\end{figure}

The experimental setup to observe superbunching effect is the same as the one in Ref. \cite{zhou-2019} except a computer  is employed to generate random numbers following binary distribution. The scheme in Fig. \ref{1} can be divided into two parts. Figure \ref{1}(a) is employed to generate the laser light with designed intensity modulation and Fig. \ref{1}(b) consists of a typical pseudothermal light source and a HBT interferometer \cite{martienssen,hbt,hbt-1}. A single-mode continuous-wave laser light is attenuated by variable attenuator (VA$_1$) before entering the intensity modulator, which consists of two orthogonally polarized polarizers (P$_1$ and P$_2$), and an EOM. The EOM is driven by a high-voltage amplifier (HV), which amplifies the voltage signals generated by a signal generator (SG) and a computer (PC). An oscilloscope (OS) is employed to monitor the voltage signals from HV and the modulated intensity via photo-detector, D$_I$. BS is a 1:1 non-polarizing beam splitter. A focus lens (L) and a rotating groundglass (RG) are employed to scatter the modulated laser light. The HBT interferometer consists of a non-polarizing 1:1 fiber beam splitter (FBS), two single-photon detectors (D$_1$ and D$_2$), and a two-photon coincidence counting detection system (CC). Similar experimental setup are employed to measure the third-order temporal coherence function by replacing FBS with a 1:1:1 fiber beam splitter and adding one more single-photon detector.

\begin{figure}[htbp]
\centering
\includegraphics[width=80mm]{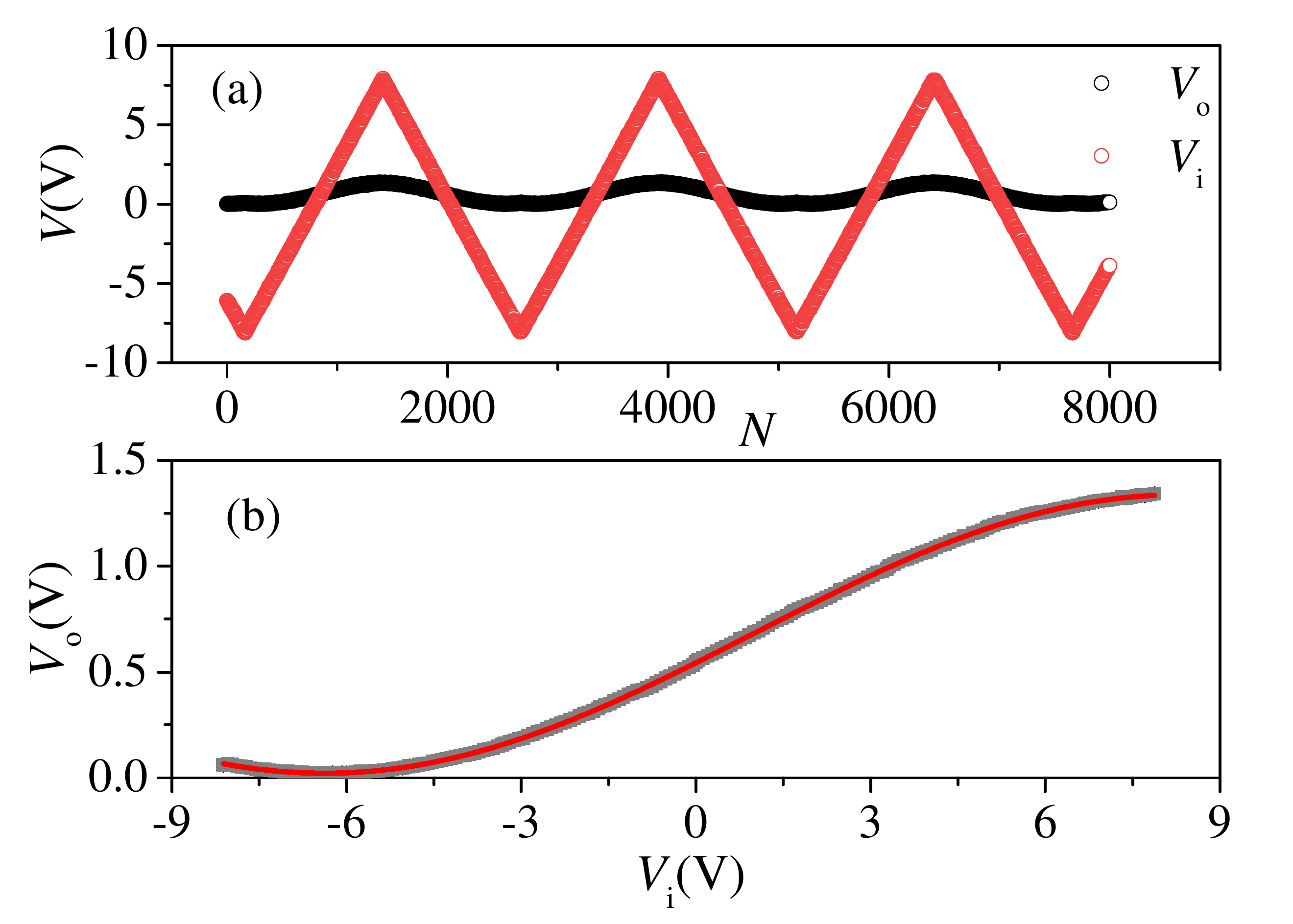}
\caption{Relation between the applied voltage from SG and output voltage of the monitoring detector. $V_{\text{o}}$: output voltage from D$_I$. $V_{\text{i}}$: input voltage from SG. $N$: series number. (a) is directly recorded by OS and (b) is the fitted relation between $V_{\text{o}}$ and $V_{\text{i}}$. }\label{2}
\end{figure}

Figure \ref{2} shows the relationship between the applied voltage from SG and output voltage of the monitoring detector. $V_{\text{o}}$ is output voltages of D$_I$ and $V_{\text{i}}$ is input voltages from signal generator. The results in Fig. \ref{2}(a) are recorded directly by the oscilloscope. The gray squares in Fig. \ref{2}(b) are obtained by summing the measured results with identical $V_{\text{i}}$, in which the error bars are too small to observe. The red line in Fig. \ref{2}(b) is theoretical fitting of the experimental data and the fitted equation is as follows,
\begin{equation}\label{fitting}
V_{\text{o}}=1.316\times \sin^2[0.107\times(V_{\text{i}}+6.378)]+0.022.
\end{equation}
With the help of Eq. (\ref{fitting}), we can obtain the maximal and minimal output voltages from D$_I$, which corresponds to the maximal and minimal modulated intensities in experiments, respectively. Letting $V_{\text{o}}$ equals the maximal and minimal values, the corresponding applied voltages can be calculated, respectively. Hence the designed intensities with two different values can be obtained by applying the calculated voltages, respectively. In the following experiments, the maximal and minimal output voltages are 1.338 and 0.022 V, in which the applied voltage from SG are calculated to be 8.30 and -6.38 V, respectively. 

\begin{figure}[htbp]
\centering
\includegraphics[width=80mm]{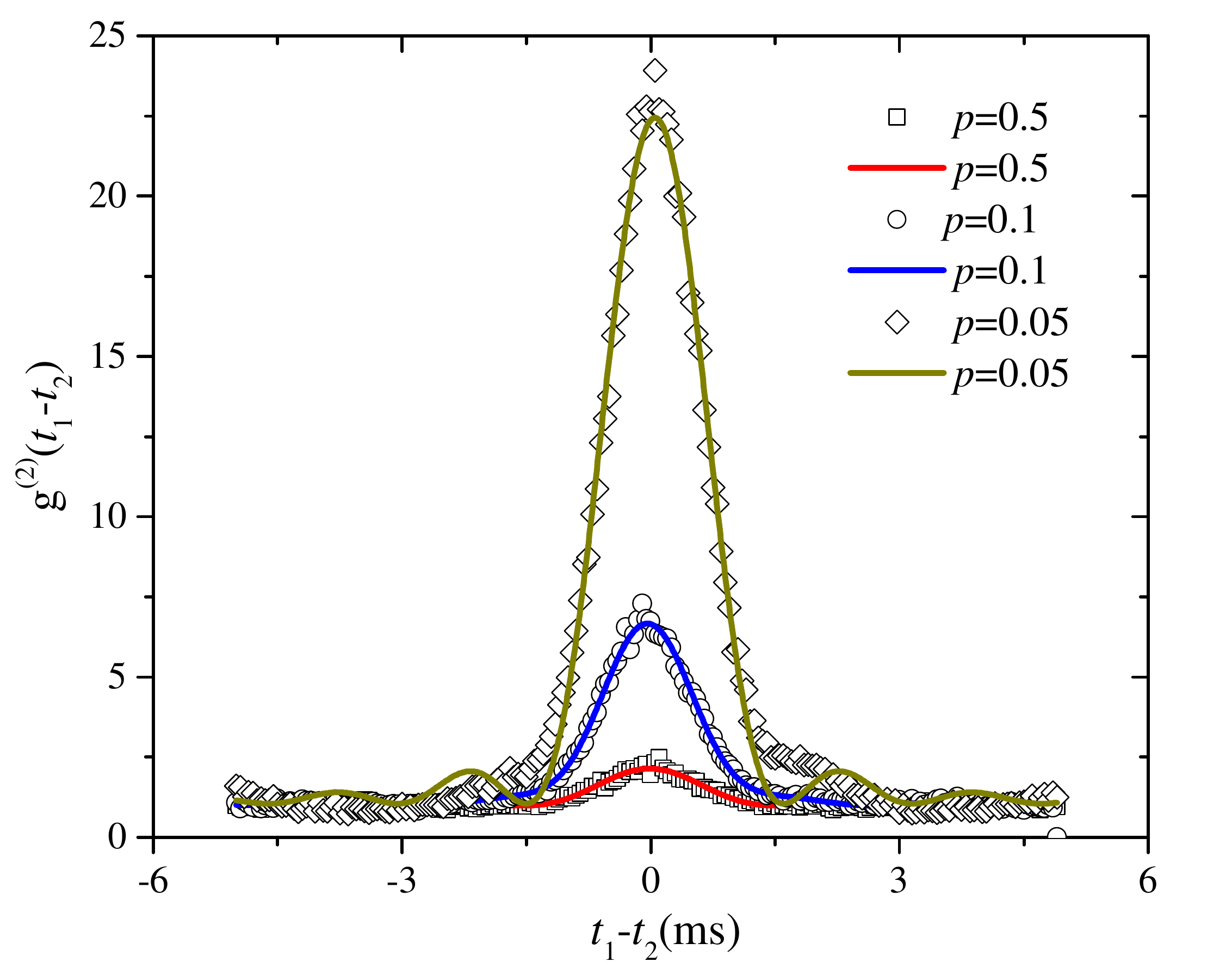}
\caption{Measured second-order temporal coherence functions of superbunching pseudothermal light. $g^{(2)}(t_1-t_2)$: normalized second-order temporal coherence function. $t_1-t_2$: time difference between the two single-photon detection events. $p$: the probability of the intensity of the incident laser light equaling 1.338 $a.u.$. The empty squares, circles, and diamonds are the experimental measured results for $p$ equaling 0.5, 0.1 and 0.05, respectively. The solid lines are theoretical fitting of the data.}\label{3}
\end{figure}

The measured second-order temporal coherence functions are shown in Fig. \ref{3}. The employed two different intensities in binary distribution are 1.338 and 0.022 $a.u$. $g^{(2)}(t_1-t_2)$ is the normalized second-order temporal coherence function. $t_1-t_2$ is the time difference between two single-photon detection events within a two-photon coincidence count. $p$ is the probability of the intensity of the incident laser light equaling 1.338 $a.u.$ in binary distribution.  The empty squares, circles, and diamonds are the experimentally measured results for $p$ equaling 0.5, 0.1 and 0.05, respectively. The solid lines are theoretical fittings of the experimental data by employing Eq. (\ref{G2t}). As $p$ decreases from 0.5 to 0.05, the degree of second-order coherence of superbunching pseudothermal light increases. For instance, $g^{(2)}(0)$ equals 2.16 when $p$ equals 0.5. When $p$ equals 0.1, $g^{(2)}(0)$ increases to 6.74. As  $p$ decreases to 0.05, $g^{(2)}(0)$ increases to 20.45. It is predicted as the value of $p$ continues to decrease, the value of $g^{(2)}(0)$ can be further increased.

\begin{figure}[htbp]
\centering
\includegraphics[width=100mm]{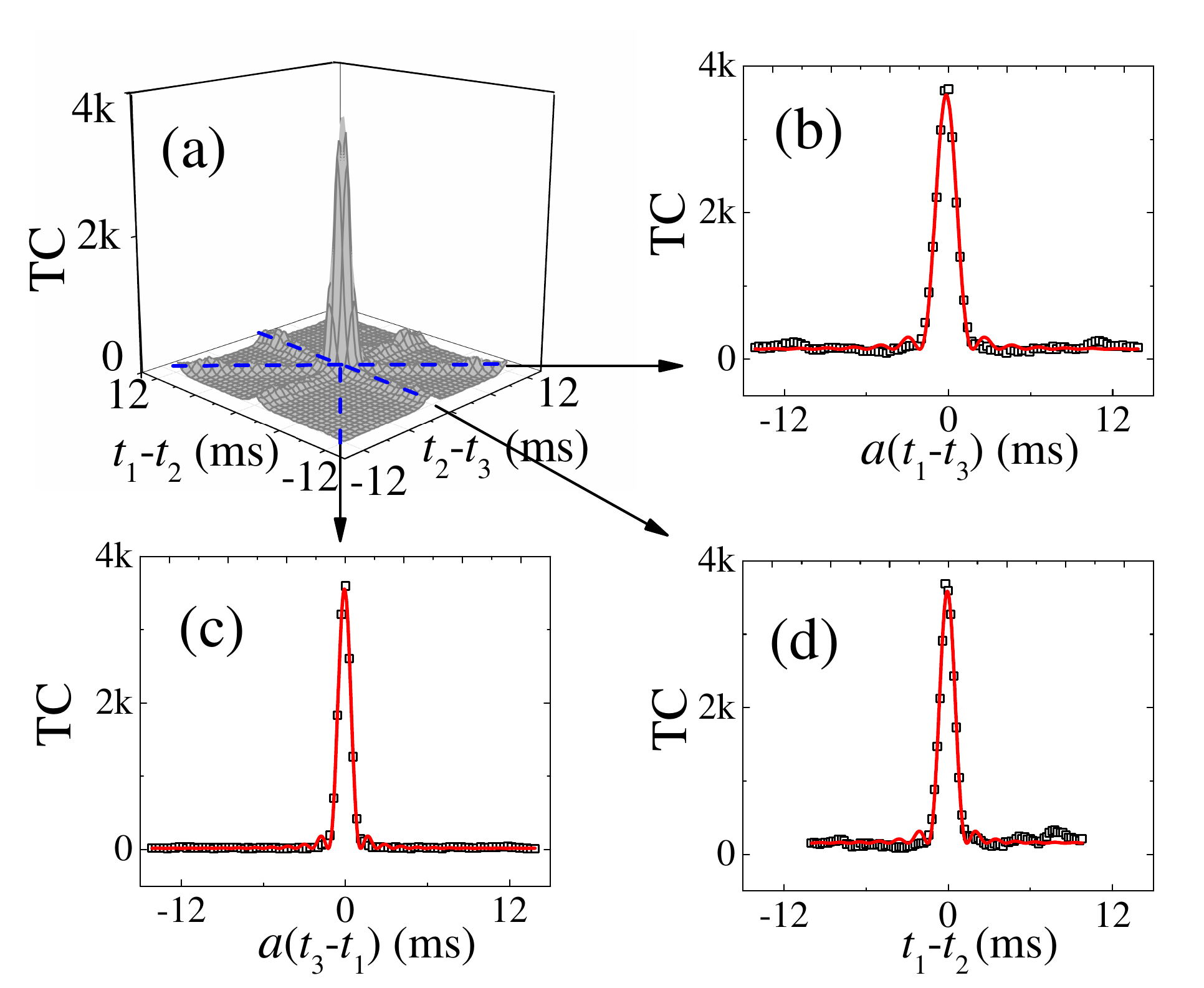}
\caption{Measured three-photon coincidence counts of superbunching pseudothermal light when $p=0.05$. TC: three-photon coincidence count. $t_1$, $t_2$, and $t_3$ are the time for photon detection events at D$_1$, D$_2$, and D$_3$, respectively. $a$ is a constant equaling $\sqrt{2}$. (b) - (d) are the cross sections of the measured TC in (a) along different directions as indicated by the black arrows, respectively.}\label{4}
\end{figure}

Figure \ref{4}(a) shows the measured third-order temporal coherence function of superbunching pseudothermal light when $p$ equals 0.05, in which TC is three-photo coincidence count, $a$ is a constant equaling $\sqrt{2}$, $t_1$, $t_2$, and $t_3$ are the time for photon detection events at D$_1$, D$_2$, and D$_3$, respectively. Figures \ref{4}(b) and \ref{4}(c) are the cross sections of the measured results in Fig. \ref{4}(a) along the directions of $t_1-t_2=-(t_2-t_3)$ and $t_1-t_2=t_2-t_3$, respectively. Figure \ref{4}(d) is the cross section of the one in Fig. \ref{4}(a) along the direction of $t_2=t_3$. The empty squares are experimental data and solid lines are theoretical fittings  \cite{zhou-josab}. The ratio between the peak and background of the measured TC in Fig. \ref{4}(c) can be treated as the measured degree of third-order coherence, $g^{(3)}(0)$, which is calculated to be 227.07.

\section{Discussions}\label{discussion}

In Sect. \ref{experiments}, we have observed $g^{(2)}(0)$ equaling 20.45 and $g^{(3)}(0)$ equaling 227.07, which is much larger than the ones of thermal light, 2 and 6, respectively. Hence two- and three-photon superbunching is observed in our experiments \cite{ficek-book}. As shown in Eq. (\ref{G2t}), the value of $g^{(2)}(0)$ can be further increased if larger correlation of intensity modulation was employed. Here, random numbers following binary distribution are employed to show how $g^{(2)}(0)$ can be changed by varying the corresponding parameters. 

\begin{figure}[htbp]
\centering
\includegraphics[width=80mm]{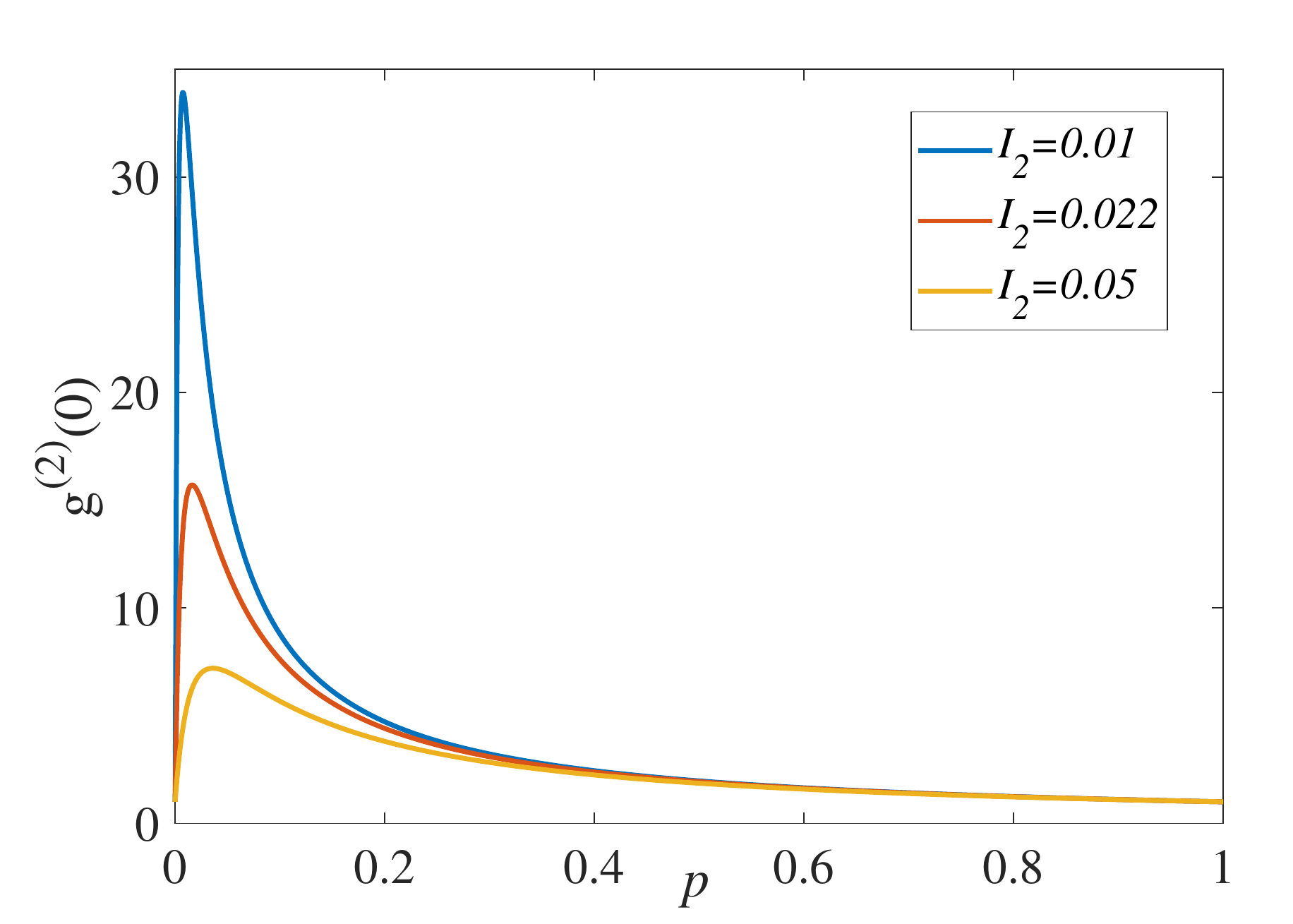}
\caption{The degree of second-order coherence of light following binary distribution with different parameters. $g^{(2)}(0)$ is the degree of second-order coherence and $p$ the probability of intensity equaling $I_1$. $I_1$ equals 1.338 $a.u.$. Three cases are for $I_2$ equaling 0.01, 0.022, and 0.05 $a.u.$, respectively. }\label{5}
\end{figure}

There are two different intensities, $I_1$ and $I_2$ for binary distribution. The probabilities of the intensity equaling $I_1$ and $I_2$ are $p$ and $1-p$, respectively. The degree of second-order coherence is defined as \cite{loudon-book}
\begin{equation}\label{g20}
g^{(2)}(0)=\frac{\langle I^2 \rangle }{\langle I \rangle^2},
\end{equation}
where $\langle... \rangle$ means ensemble average. Substituting the above parameters for binary distribution into Eq. (\ref{g20}), it is straightforward to have 
\begin{equation}\label{g21}
g^{(2)}(0)=\frac{pI_1^2+(1-p)I_2^2}{[pI_1+(1-p)I_2]^2}=\frac{pR^2+(1-p)}{[pR+(1-p)]^2}, 
\end{equation}
where $R$ is defined as $I_1/I_2$. The value of $g^{(2)}(0)$ is dependent on both the values of $p$ and $R$. Here we will discuss the situation for $I_1$ and $I_2$ equaling 1.338 and 0.022 $a.u$, respectively, which are the same as the ones in our experiments. The red line in Fig. \ref{5} shows how the value of $g^{(2)}(0)$ changes with $p$. The maximal value of $g^{(2)}(0)$ is 15.71, which is obtained when $p$ equals 0.016. When $p$ equals 0.05,  $g^{(2)}(0)$ equals 11.67. It is consistent with the experimental results, 20.45, if the effect of rotating groundglass on $g^{(2)}(0)$ was considered by employing Eq. (\ref{G2t}). In order to show how to further increase the value of $g^{(2)}(0)$, we also plot the cases for $I_2$ equaling 0.01 and 0.05 in Fig. \ref{5}. The maximal value of $g^{(2)}(0)$ will increase as the ratio between $I_1$ and $I_2$ increases in the binary distribution.

\begin{figure}[htbp]
\centering
\includegraphics[width=80mm]{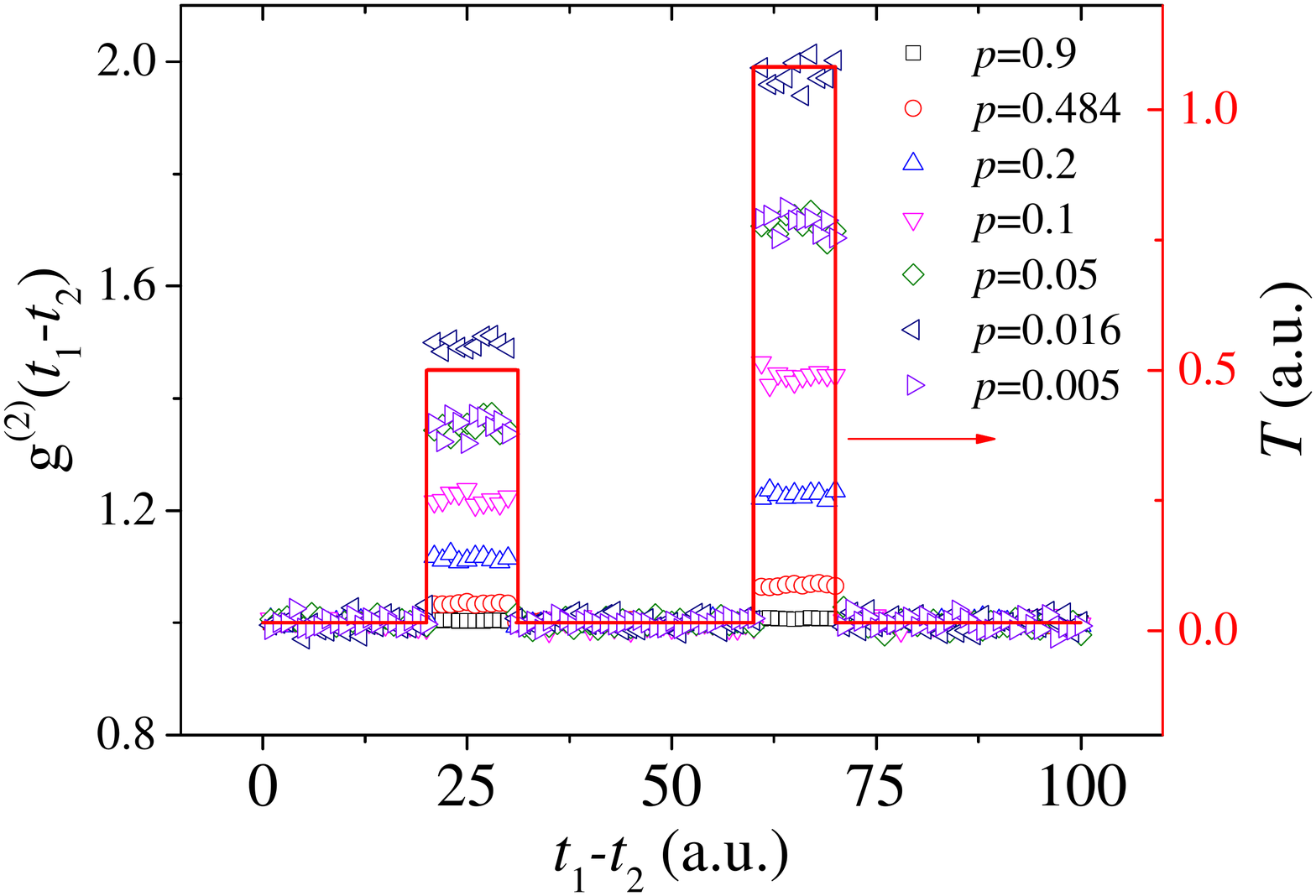}
\caption{Simulated temporal ghost imaging with random intensities following binary distribution. $I_1$ and $I_2$ equals 1.338 and 0.022 $a.u$, respectively.  }\label{6}
\end{figure}

Figure \ref{6} shows the simulated temporal ghost imaging with random intensities following binary distribution. The parameters employed in the simulation are as follows. $I_1$ and $I_2$ equals 1.338 and 0.022 $a.u$, respectively. $p$ is the probability of equaling $I_1$. The employed temporal object is a double pulses with different heights as the one shown by the solid line in Fig. \ref{6}, which can be experimentally implemented by an EOM \cite{tgi-2016}. There are three different values for the temporal object, 0, 0.5, and 1, which correspond to 0, 50\%, and 100\% of the input signal passing through, respectively. The symbols are the numerical simulations of temporal ghost imaging when $p$ equals 0.9, 0.484, 0.2, 0.1, 0.05, 0.016, and 0.005. The reason why $p=0.484$ is chosen is that $g^{(2)}(0)$ equaling 2 in this condition, which is the same as the one of thermal or pseudothermal light and can be treated as a reference. The reason why $p=0.016$ is chosen is that  $g^{(2)}(0)$  gets its maximal value, 15.71, in this condition.

The visibility (Vis) and signal-to-noise ratio (SNR) of the retrieved images for different values of $p$ are listed in table \ref{t1}. The definition for visibility is \cite{liu-2018}
\begin{equation}
\text{Vis}=\frac{g^{(2)}_{\text{max}}-1}{g^{(2)}_{\text{max}}+1},
\end{equation}
in which $g^{(2)}_{\text{max}}$ is the average value of the retrieved temporal ghost imaging for all the points with transmission equaling 1, \textit{i.e.}, the average value of the points on the top of the second pulse in Fig.  \ref{6}. The employed definition for calculating SNR is \cite{snr,snr-book}
\begin{equation}
\text{SNR}=\frac{\sum_j[T(j)-\bar{T}]^2}{\sum_j[g^{(2)}(j)-T(j)]^2},
\end{equation}
where the meanings of $T(j)$ and $g^{(2)}(j)$ are the same as the ones in Fig. \ref{6}. $\bar{T}$ is the average of $T$.

\begin{table}[htb]
\centering
\begin{tabular}[c]{|c|c|c|c|c|c|c|c|}
\hline
$p$&0.9&0.484&0.2&0.1&0.05&0.016&0.005\\
\hline
Vis(\%)&0.34&3.26&10.26&18.06&26.16&32.82&26.25\\
\hline
SNR&0.1242&0.1237&0.1216&0.1176&0.1111&0.1030&0.1110\\
\hline
\end{tabular}
\caption{Visibility and signal-to-noise ratio of simulated temporal ghost imaging with random numbers following binary distribution for different values of $p$. Vis: visibility. SNR: signal-to-noise ratio.}\label{t1}
\end{table}

As $p$ increases from 0.484 to 0.016, the visibility of the retrieved image increases about 10 times, \textit{i.e.}, from 3.26\% to 32.82\%. At the same time, the SNR decreases from 0.1237 to 0.1030, which decreases about 16.73\%. The simulation confirms that the SNR of temporal ghost imaging with binary distribution does not decrease dramatically when the visibility increases, which indicates that this type of superbunching pseudothermal light can be employed to improve the image quality of temporal ghost imaging. 

\section{Conclusions}\label{conclusion}

In conclusions, we have proposed a simple and efficient method to generate superbunching pseudothermal light with tunable degree of second-order coherence. We experimentally observed $g^{(2)}(0)$ and $g^{(3)}(0)$ equaling 20.45 and 227.07, respectively, which are much larger than the ones of thermal or pseudothermal light.  Unlike the one with multiple rotating groundglasses \cite{zhou-2017,liu-2018}, this newly proposed superbunching psedothermal light source can be employed to increase the visibility of temporal ghost imaging, while keeping the signal-to-noise ratio almost unchanged. By changing the parameters of binary distribution, the visibility of temporal ghost imaging can be increased from 3.26\% to 32.82\%, while the signal-to-noise ratio  decreases from 0.1237 to 0.1030. The newly proposed superbunching pseudothermal light source can be applied in temporal ghost imaging, multi-photon interference, and other applications in quantum optics where thermal or pseudothermal light source were needed.

Another interesting point worthy of noticing is that the difference between ghost imaging with classical light and quantum light. It is concluded that ghost imaging with thermal light can mimic all the aspects of ghost imaging with entangled photon pairs except lower visibility \cite{gi-2004}. By employing superbunching pseudothermal light, the difference may no longer exist. It would be beneficial to understand the difference between ghost imaging with quantum and classical light by taking superbunching pseudothermal light into account.

\section*{Acknowledgments}
This project is supported by Shanxi Key Research and Development Project (Grant No. 2019ZDLGY09-08), Open fund of MOE Key Laboratory of Weak-Light Nonlinear Photonics (OS19-2), and the Fundamental Research Funds for the Central Universities.

\bibliography{sample}


\end{document}